\documentclass[twocolumn,floatfix,superscriptaddress,showpacs,prb]{revtex4}
\usepackage{graphicx}
\usepackage{amsmath}
\usepackage{bm}
\begin{document}

\title{Reentrance effect in a graphene \textit{n-p-n} junction coupled to a superconductor}
\author{A. Ossipov}
\affiliation{Instituut-Lorentz, Universiteit Leiden, P.O. Box 9506, 2300 RA Leiden, The Netherlands}
\author{M. Titov}
\affiliation{Department of Physics, Konstanz University, D--78457 Konstanz, Germany}
\author{C. W. J. Beenakker}
\affiliation{Instituut-Lorentz, Universiteit Leiden, P.O. Box 9506, 2300 RA Leiden, The Netherlands}

\date{April 2006}
\begin{abstract}
We study the interplay of Klein tunneling (= interband tunneling) between \textit{n-}doped and \textit{p-}doped regions in graphene and Andreev reflection (= electron-hole conversion) at a superconducting electrode. The tunneling conductance of an \textit{n-p-n} junction initially increases upon lowering the temperature, while the coherence time of the electron-hole pairs is still less than their lifetime, but then drops back again when the coherence time exceeds the lifetime. This reentrance effect, known from diffusive conductors and ballistic quantum dots, provides a method to detect phase coherent Klein tunneling of electron-hole pairs.
\end{abstract}
\maketitle

The conductance of a diffusive conductor increases if one of the electrodes becomes superconducting upon lowering the temperature, as a consequence of phase coherence between electron and hole excitations (Andreev pairs) induced by the proximity to the superconductor. The initial increase does not persist to the lowest temperatures. Instead, the normal-state value reappears when the thermal coherence length of the Andreev pairs exceeds the sample size. This is the so called reentrance effect, first observed a decade ago in a metal wire\cite{Cha96,Chi99} and in a semiconductor two-dimensional electron gas.\cite{Har96,Toy99} (The theoretical prediction goes back much further.\cite{Naz96,Art79}) The reentrance effect is now routinely used to measure the coherence time of Andreev pairs in a much wider class of diffusive conductors, see for example a recent experiment on multiwall carbon nanotubes.\cite{Har03} It has also been predicted to occur in a ballistic quantum dot with properly adjusted point contacts.\cite{Cle00} We refer to Ref.\ \onlinecite{Cou99} for an extensive review of the topic and to Ref.\ \onlinecite{Bee00} for a tutorial.

As pointed out by Silvestrov and Efetov,\cite{Sil07} a \textit{p-}doped region in \textit{n-}doped graphene (or, converselyly, an \textit{n-}doped region in \textit{p}-doped graphene) confines carriers in much the same way as a quantum dot in a two-dimensional electron gas. An electron in the valence band of the \textit{p-}doped region can escape into the conduction band of the adjacent \textit{n-}doped regions, by the process of Klein tunneling.\cite{Che06,Kat06} This interband tunneling process is highly directional: Only electrons near normal incidence are transmitted through a smooth \textit{n-p} interface.\cite{Che06} The \textit{n-p} interface thus functions as a constriction in momentum space, analogously to the constriction in real space formed by a point contact in a conventional quantum dot. 

If an \textit{n-p} junction is in series with a superconducting electrode, then the interband tunneling is combined with electron-hole conversion (known as Andreev reflection\cite{And64}) at the interface with the superconductor. Here we study the interplay of these two scattering mechanisms, and show that they lead to a reentrance effect in the temperature dependent conductance.

\begin{figure}[tb]
\bigskip

\includegraphics[width=0.9\linewidth]{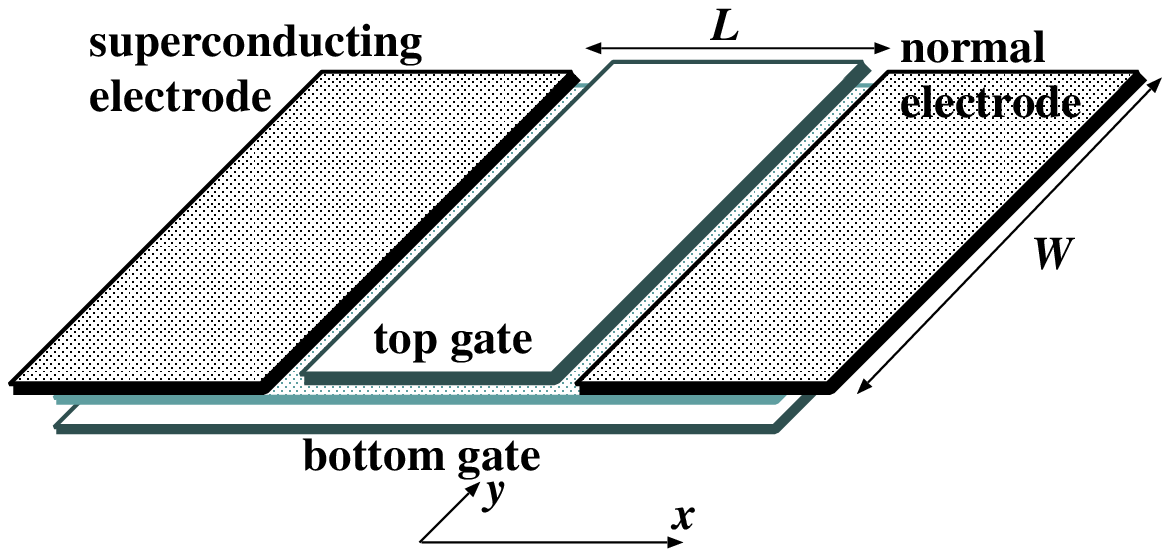}

\includegraphics[width=0.7\linewidth]{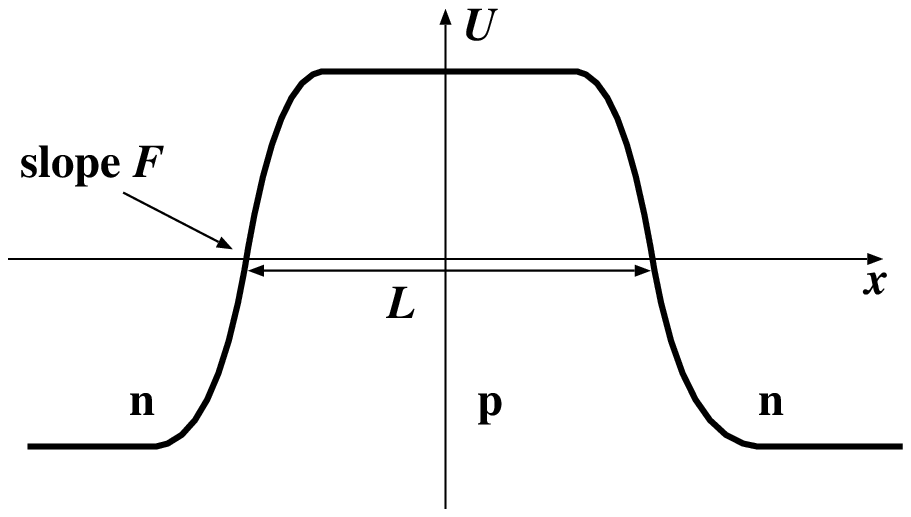}
\caption{\label{fig_SnpnN}
Top panel: Graphene layer connected to one normal metal electrode and one superconducting electrode. A bottom and top gate control the electrostatic doping, so that the graphene regions covered by the electrodes are \textit{n-}type and the central region is  \textit{p}-type. Bottom panel: electrostatic potential profile across the \textit{n-p-n} junction. The Fermi level (at $E_{F}=0$) lies in the conduction band in the two \textit{n-}regions, while it lies in the valence band in the \textit{p}-region.
}
\end{figure}

We consider a \textit{p-}doped graphene strip (length $L$, width $W$), connecting two \textit{n-}doped regions (see Fig.\ \ref{fig_SnpnN}). One \textit{n-}region is contacted by a normal metal electrode, the other by a superconducting electrode. The electrostatic potential profile, controlled by a top gate and a bottom gate, is assumed to vary smoothly on the scale of the Fermi wave length. Close to the Fermi level (chosen at $E_{F}=0$) the potential energy increases as $U(x)=F(x+L/2)$ near the left \textit{n-p} interface and decreases as $U(x)=-F'(x-L/2)$ near the right \textit{p-n} interface. The slopes $F,F'$ are of order $\hbar v\bar{k}_{F}/d$, with $v$ the carrier velocity in graphene, $\bar{k}_{F}$ the average of the Fermi wave vector at the two sides of the interface, and $d$ the thickness of the interface. For simplicity, we will take $F=F'$ in the main text and consider the effects of two unequal interfaces at the end of the paper.

The tunnel probabilities $T_{n}$ per mode through a smooth \textit{p-n}  junction ($\bar{k}_{F}d\gg 1$) have been calculated by Cheianov and Fal'ko:\cite{Che06}
\begin{equation}
T_{n}=\exp(-\pi\hbar v q_{n}^{2}/F),\label{Tnresult}
\end{equation}
in terms of the transverse wave vector $q_{n}$ of the $n$-th mode. These determine the tunnel conductance via the Landauer formula,
\begin{equation}
G_{\textit{p-n}}=g_{0}\sum_{n}T_{n},\;\;g_{0}=4e^{2}/h.\label{Landauer}
\end{equation}
The total number of propagating modes in the \textit{p-}doped region is $N= k_{F}W/\pi$, with $k_{F}$ the Fermi wave vector in that region. Only a relatively small number $N'\simeq W(\bar{k}_{F}/d)^{1/2}$ of these modes near normal incidence have $T_{n}$ close to 1. (In this sense a \textit{p-n} interface forms a constriction in momentum space.) We assume that the number $N'$ of open scattering channels is still $\gg 1$, so that sums over $n$ may be replaced by integrations over $q$: $\sum_{n}\rightarrow (W/\pi)\int_{0}^{\infty}\,dq$. The resulting tunnel conductance is\cite{Che06}
\begin{equation}
G_{\textit{p-n}}=\frac{g_{0}W}{2\pi}\sqrt{\frac{F}{\hbar v}}.\label{Gpnresult}
\end{equation}

Eq.\ \eqref{Gpnresult} is the conductance of a \textit{p-n} junction between two normal metal contacts. If one of the contacts is superconducting we can calculate the conductance from the formula\cite{Akh07}
\begin{equation}
G_{\textit{p-n}}^{\rm A}=g_{0}\sum_{n}\frac{2T_{n}^{2}}{(2-T_{n})^{2}}.\label{GpnA}
\end{equation}
Replacing again the sum over modes by an integration, this Andreev conductance evaluates to
\begin{equation}
G_{\textit{p-n}}^{\rm A}=1.082\,G_{\textit{p-n}}.\label{GpnAresult}
\end{equation}
The incoherent series conductance $G_{\rm incoherent}$ of the \textit{n-p-n} junction between a superconducting and normal metal contact becomes
\begin{equation}
G_{\rm incoherent}=\frac{G_{\textit{p-n}}G_{\textit{p-n}}^{\rm A}}{G_{\textit{p-n}}+G_{\textit{p-n}}^{\rm A}}=1.040\,G_{0}.\label{Gincoh}
\end{equation}
It is slightly larger than the series conductance $G_{0}=\frac{1}{2}G_{\textit{p-n}}$ when both contacts are in the normal state.

To determine the coherent series conductance we assume that the \textit{p}-region is weakly disordered (mean free path $l\lesssim L$), such that the modes are randomized before the electron or hole escapes out of the \textit{p}-doped region. We thus require that the scattering time $\tau=l/v$ is less than the dwell time $\tau_{\rm dwell}\simeq (L/v)(N/N')$, which is satisfied if $\sqrt{k_{F}d}\gg 1$ (assuming $k_{F}\simeq \bar{k}_{F}$). During a time $\tau_{\rm dwell}$ the carrier explores an area $LW_{\rm eff}$, with $W_{\rm eff}=\min(W,\sqrt{D\tau_{\rm dwell}})$ determined by the diffusion constant $D\simeq vl$. The \textit{n-p-n} junction corresponds statistically to $W/W_{\rm eff}$ quantum dots in parallel, each with $N_{\rm eff}=(W_{\rm eff}/W)N'$ open scattering channels. For $N_{\rm eff}\gg 1$ we may use random-matrix theory to calculate the average density of transmission eigenvalues $\rho(T)$ through this system. The difference between the symplectic ensemble appropriate for Dirac fermions in graphene and the orthogonal ensemble of a conventional quantum dot does not show up to leading order in $N_{\rm eff}$, so we may ignore it and use the general result in the literature for the orthogonal ensemble,\cite{Bro96} 
\begin{equation}
\rho(T)=\frac{1}{\pi\sqrt{T(1-T)}}\sum_{n}\frac{T_{n}(2-T_{n})}{T_{n}^{2}-4T_{n}T+4T}.\label{rhoTresult}
\end{equation}
Combining Eqs.\ \eqref{Tnresult} and \eqref{rhoTresult}, the coherent series conductance $G_{\rm coherent}$ evaluates to
\begin{equation}
G_{\rm coherent}=g_{0}\int_{0}^{1}dT\,\rho(T)\frac{2T^{2}}{(2-T)^{2}}=1.019\,G_{0}.\label{Gcoherent}
\end{equation}

The crossover from coherent to incoherent series addition occurs when the thermal energy $kT$ becomes comparable to the Thouless energy
\begin{equation}
E_{T}=\frac{G_{\textit{p-n}}\delta}{2\pi g_{0}}=\frac{\sqrt{\hbar vF}}{2\pi k_{F}L} ,\label{ETdef}
\end{equation}
with $\delta=2\pi\hbar v/(LWk_{F})$ the mean level spacing (per spin and valley) in the \textit{p}-region. In order of magnitude, $E_{T}\simeq (\hbar v/L)(k_{F}d)^{-1/2}\simeq\hbar/\tau_{\rm dwell}$. We assume that the gap $\Delta$ in the superconducting reservoir is $\gg E_{T}$. From Eqs.\ \eqref{Gincoh} and \eqref{Gcoherent} we expect a 2\% decrease of the conductance upon lowering the temperature $T$ below $E_{T}/k$. We will now show that this decrease is preceded by an increase, such that the conductance is maximal at $kT\simeq E_{T}$.

To calculate this reentrance effect we may again use random-matrix theory, as in Ref.\ \onlinecite{Cle00}, or we may use the equivalent circuit theory of quasiclassical Green's functions,\cite{Naz99} as in Refs.\ \onlinecite{Sam04,Big06}. Here we follow the latter approach. Just as in the random-matrix approach, no modification of the conventional circuit theory is needed to leading order in $N_{\rm eff}$. In this quasiclassical limit the Green's functions $\check{G}(\bm{r},\bm{n})$ are represented by $4\times 4$ matrices acting in the Keldysh and Nambu spaces, as a function of position $\bm{r}$ and momentum direction $\bm{n}$. Green's functions $\cal{G}$ for Dirac fermions have an additional $\rm{SU}(2)\times\rm{SU}(2)$ structure from the valley and pseudospin degree of freedom,\cite{Khv06,McC06} which in the quasiclassical limit factors out:
\begin{equation}
{\cal G}(\bm{r},\bm{n})=\tfrac{1}{2}\check{G}(\bm{r},\bm{n})\otimes\left(1+n_{x}\sigma_{1}+n_{y}\sigma_{2}\right)\otimes\tau_{0}.\label{calGdef}
\end{equation}
[The Pauli matrices $\tau_{i}$ and $\sigma_{i}$ act on the valley and pseudospin degree of freedom, respectively, and we take a basis in which the Dirac Hamiltonian is $v(\bm{p}\cdot\bm{\sigma})\otimes\tau_{0}$.]

The quasiclassical Green's functions in the superconducting lead $\check{G}_S$, in the normal lead $\check{G}_N$, and in the \textit{n-p-n} junction $\check{G}$ are $2\times 2$ matrices in the Keldysh space,\cite{Ram86}
\begin{equation}
\check{G}=\begin{pmatrix} 
\hat{R}&\hat{K} \\
0&\hat{A}
\end{pmatrix},\label{GRKAdef}
\end{equation}
where $\hat{R},\hat{A}$, and $\hat{K}$ are $2\times 2$ matrices in the Nambu space. In the leads they take their equilibrium values, depending only on the excitation energy $\varepsilon$:
\begin{align}\label{normal}
\hat{R}_N&=-\hat{A}_N=\Lambda_{3},\;\;\ \hat{K}_N=2[f_{-}+(1-f_{+})\Lambda_{3}],\\
\label{super} 
\hat{R}_S&=\hat{A}_S=\Lambda_{1},\;\;\hat{K}_S=0.
\end{align}
Here $f_{\pm}=f(\varepsilon-eV)\pm f(\varepsilon+eV)$, with $V$ the applied voltage and $f(\varepsilon)=(1+e^{\varepsilon/kT})^{-1}$ the Fermi function. The Pauli matrices $\Lambda_{i}$ act in the Nambu space.

The Green's function in the \textit{n-p-n} junction has the form
\begin{align}
\hat{R}&=\Lambda_{1}\cos\alpha+\Lambda_{3}\sin\alpha,\;\;
\hat{A}=\Lambda_{1}\cos\alpha^\ast-\Lambda_{3}\sin\alpha^\ast,\nonumber\\
\hat{K}&=\hat{R}\hat{F}-\hat{F}\hat{A},\;\;
\hat{F}=1-F_{+}+F_{-}\Lambda_{3}.
\end{align}
The unknown parameters $\alpha$, $F_{\pm}$ can be found from the
conservation law of the matrix current,\cite{Naz99}
\begin{equation}\label{conserv}
\check{I}_{N}+\check{I}_{S}+\check{I}_{\varepsilon}=0,
\end{equation}
with the definitions
\begin{equation}\label{m-current}
\check{I}_{X}=\sum_n\frac{2g_{0} T_n [\check{G}_X,\check{G}]}
{4+T_n(\{\check{G}_X,\check{G}\}-2)},\;\;
\check{I}_{\varepsilon}=\frac{2\pi i g_0\varepsilon}{\delta} [\check{G},\Lambda_{3}].
\end{equation}
Here $[\cdots]$ is a commutator, $\{\cdots\}$ is an anticommutator, and the label $X$ stands for $N$ or $S$.

From the retarded component of Eq.\ \eqref{conserv} we obtain an
equation for $\alpha$,
\begin{align}\label{alpha}
&\tan\alpha\,Z(\cos\alpha)=Z(\sin\alpha)-
\frac{2\pi i\varepsilon}{\delta},\\
&Z(x)=\sum_n(2T_n^{-1}+x-1)^{-1}.
\end{align}
The Keldysh component of Eq.\ \eqref{conserv} determines $F_{\pm}$,
\begin{equation}\label{keldysh}
F_{+}=f_{+},\;\;
F_{-}=\frac{f_{-}\,{\rm Im}\,[\cos\alpha\,Z(\sin\alpha)]}
{{\rm Im}\,[\cos\alpha\,Z(\sin\alpha)-\sin\alpha\,Z(\cos\alpha)]}.
\end{equation}
The electrical current $I$ is related to the Keldysh component $(\check{I}_{N})_K$ of the matrix current,
\begin{equation}
I=-\frac{1}{2e}\int_0^{\infty}d\varepsilon\,{\rm Tr}\,[\Lambda_{3} (\check{I}_{N})_K].
\end{equation} 
The zero temperature differential conductance $G(V)=dI/dV$ is then given by  
\begin{equation}\label{cond-tot}
G(V)=2g_0(F_{-}/f_{-})\coth ({\rm Im}\,\alpha)
\,{\rm Im}\,[\sin\alpha\,Z(\cos\alpha)],
\end{equation}
where $\alpha$ is a solution of Eq.\ \eqref{alpha} at $\varepsilon=eV$ and $F_{-}/f_{-}$ is given 
by Eq.\ \eqref{keldysh}.

\begin{figure}[tb]
\bigskip
\includegraphics[width=0.9\linewidth]{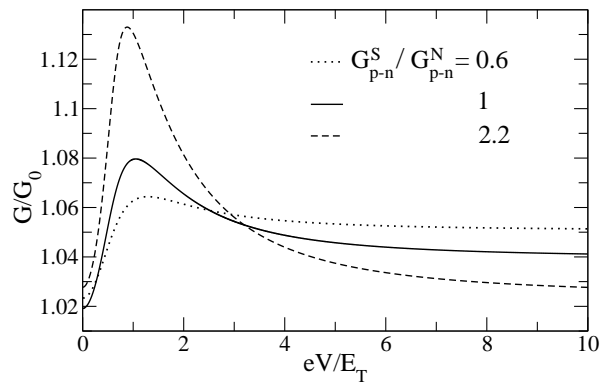}
\caption{\label{G_V} Differential conductance normalized to its normal-state value as a function of the applied voltage, for three values of the ratio $G_{\textit{p-n}}^{S}/G_{\textit{p-n}}^{N}$ (with $G_{\textit{p-n}}^{S}$ the tunnel conductance of the \textit{p-n} junction closest to the superconductor and $G_{\textit{p-n}}^{N}$ the tunnel conductance of the other interface). The normal-state conductance $G_{0}$ is the series conductance defined by $1/G_{0}=1/G_{\textit{p-n}}^{N}+1/G_{\textit{p-n}}^{S}$.}
\end{figure}

\begin{figure}[tb]
\includegraphics[width=0.9\linewidth]{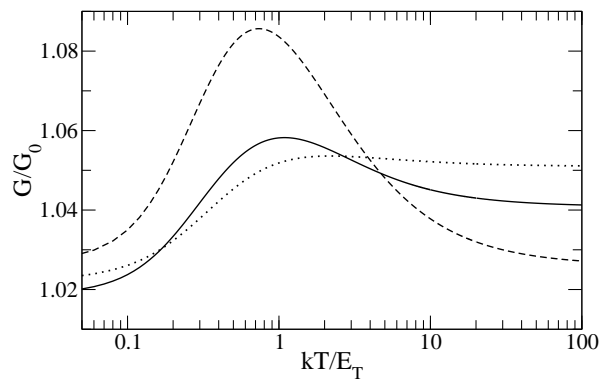}
\caption{\label{G_T} Same as Fig.\ \ref{G_V}, but now for the linear response conductance as a function of temperature.}
\end{figure}

The result for $G(V)$ is presented by a solid line in Fig.\ \ref{G_V}. In the limits $V\rightarrow \infty$ and $V\rightarrow 0$ we recover, respectively, the incoherent limit (\ref{Gincoh}) and the coherent limit (\ref{Gcoherent}). We find that $G(V)$ takes a maximum value $G_{\rm max}=1.080\, G_0$ at $eV\approx E_T$. Fig.\ \ref{G_T} (solid line) shows the temperature-dependent linear-response conductance
\begin{equation}
G=-2\int_0^{\infty}
d\varepsilon\, G(\varepsilon/e)\frac{d}{d \varepsilon} f(\varepsilon).
\end{equation}
Again we observe nonmonotonic behavior with a maximum value  $G_{\rm max}= 1.058\, G_0$ at $kT\approx E_T$.

Figs.\ \ref{G_V} and \ref{G_T} also show results for the more general case of two unequal \textit{p-n} interfaces, for several values of the ratio of the tunnel conductances. If the interface closest to the superconductor has a higher conductance than the other interface, then the reentrance effect is enhanced (dashed line), while it is reduced in the opposite case (dotted line). This dependence is analogous to that found in a conventional quantum dot.\cite{Cle00}

In conclusion, we have shown that the conductance of a weakly disordered \textit{n-p-n} junction in graphene coupled to a superconductor exhibits a reentrance effect, similar to what is found in diffusive conductors and ballistic quantum dots: The conductance takes approximately the same value at low and high voltages (or temperatures) and reaches a maximum when $eV$ (or $kT$) is approximately equal to the Thouless energy. This behavior should be observable in currently available graphene-superconductor junctions\cite{Hee07} and would provide a demonstration of phase-coherent Klein tunneling of electron-hole pairs.

We have benefited from discussions with A. R. Akhmerov, W. Belzig, and Yu.\ V. Nazarov. This research was supported by the Dutch Science Foundation NWO/FOM and by the German Science Foundation DFG through SFB 513.

\end{document}